\documentclass[a4paper, 11pt,english]{article}

\usepackage[T1]{fontenc}

\usepackage{amsmath}
\usepackage{amstext}
\usepackage{wasysym}
\usepackage{relsize}
\usepackage{esint}
\usepackage{babel}
\usepackage{mathtools}
\usepackage{relsize}
\usepackage{lscape}
\usepackage{cancel}
\usepackage{bm}
\usepackage{bbm}
\usepackage{multirow} 
\usepackage{amsfonts}
\usepackage{amssymb}
\usepackage{graphicx}
\usepackage[usenames,dvipsnames]{color}
\usepackage{hyperref}
\usepackage{epstopdf,epsfig}
\usepackage{verbatim}
\usepackage[utf8]{inputenc}
\usepackage[left=1in,right=1in,top=1in,bottom=1in]{geometry}             
\usepackage{booktabs} 
\usepackage{array} 
\usepackage{paralist} 
\usepackage{verbatim} 
\usepackage{subfig} 

\hypersetup{
    colorlinks=true,         
    linkcolor=blue,          
    citecolor=red,        
    urlcolor=Violet             
}

\begin{document}

\title{Connecting Holographic Wess--Zumino Consistency Condition to the Holographic Anomaly}
\author{{\bf Vasudev Shyam}\footnote{\href{mailto:vshyam@pitp.ca}{vshyam@pitp.ca}} \\\it Perimeter Institute for Theoretical Physics\\ \it 31 N. Caroline St. Waterloo, ON, N2L 2Y5, Canada }
\date{\today}                                           
\maketitle
\begin{abstract}
The Holographic Wess--Zumino (HWZ) consistency condition is shown through a step by step mapping of renormalization group flows to Hamiltonian systems, to lead to the Holographic anomaly. This condition codifies how the energy scale, when treated as the emergent bulk direction in Holographic theories, is put on equal footing as the other directions of the space the field theory inhabits. So, this is a defining feature of theories possessing local Holographic bulk duals. In four dimensional Holographic conformal field theories, the $a$ and $c$ anomaly coefficients are equated, and this is seen as a defining property of theories which possess General Relativity coupled to matter as a dual. Hence, showing how the former consistency condition leads to the latter relation between anomaly coefficients adds evidence to the claim that the HWZ condition is a defining feature of theories possessing local gravity duals. 
\end{abstract}
\tableofcontents
\section{Introduction}
A Holographic quantum field theory is one whose renormalization group flow can be mapped into a dynamical system in one higher dimension. The sources of the theory are mapped into the dynamical fields in the dual description and the energy scale or RG time is the evolution parameter for the dual fields. In order to track the renormalization group flow of space dependent sources, the local renormalization group need be employed. This is a perspective in which the couplings are upgraded into background sources and the coarse graining transformations are implemented through local Weyl rescaling of the background metric. Holographic theories are those which turn the RG flow equations of these background sources into dynamical equations of motion of fields in one higher dimension.

When the bulk theory is expected to be just classical general relativity coupled to matter fields, the relevant subset of these holographic theories are in a regime characterised by possessing a large number of degrees of freedom in which all the operators barring a finite set gain an infinitely large anomalous dimension. The few composite operators whose scaling dimensions are finite are those whose sources are the bulk fields and typically includes the metric. However, in several known examples where the holographic duality is manifest, an infinite number of single trace operators have protected scaling dimensions and therefore gain no corrections even in the strong coupling limit.\footnote{Examples include the BPS operators in 4D $\mathcal{N}=4$ SYM which are dual to KK modes in supergravity in the bulk supergravity theory on $AdS_{5}\times S_{5}$.}

Restricting attention to the case where the dynamics of the bulk theory is dictated by general relativity, to be borne out, it is necessary to ensure that diffeomorphism invariance in the bulk emerges. This means that the RG time or energy scale in the quantum field theory when mapped into a dimension of the bulk space is then treated on equal footing as the other directions of the boundary space. This concretely translates into the demand that the bulk Hamiltonian system is totally constrained as expected of one describing reparameterization and refoliation invariant systems. The feature of general covariance is encoded in a specific Poisson bracket algebra of these constraints. When translated back into field theoretic terms, this is the Holographic Wess--Zumino consistency condition. In the case where the only source to consider is the metric tensor, i.e. if for some reason the only operator that the renormalization group acts on is the energy momentum tensor in the holographic quantum field theory, this specific Poisson algebra of the constraints actually fixes the form of the constraint completely. If an additional scalar operator is also included in the set of composite operators that possess finite scaling dimension- then the Poisson algebra can fix the form of the Hamiltonian and momentum densities up to the potential for the scalar field. Constructing these Hamiltonians will be the aim of the third and fourth sections. 

In the fifth and final section, the step of renormalization is formulated. This corresponds to finding a chart on the space of sources, or `theory space'  where one can take the flow time to infinity and also define transition maps away from said chart that are reasonably well behaved. Given that the renormalization group flow is mapped into a Hamiltonian system, this chart translates into a set of local phase space co-ordinates. This is accomplished through a canonical transformation from an arbitrary set of co-ordinates to the ones where one can take the flow time to infinity. The regularity of this canonical transformation results from the aforementioned regularity in the transition maps on theory space. Given that the bulk theory of interest is gravitational and that the flow time is related to a direction in the bulk spacetime, the renormalization procedure in the field theory is related to that of finding asymptotic solutions to the bulk equations of motion. This is what is known as Holographic renormalization. Also, in this section, it is shown how the conformal algebra at infinity and consequently the asymptotic isometry group of Asymptotically Anti-deSitter spaces as duals to UV conformal field theories emerges, thereby making clear the relationship between the formalism presented in the first four sections and the AdS/CFT correspondence. 

One of the celebrated results of Holographic renormalization is the computation of the trace anomaly of the holographic quantum field theory at its ultraviolet fixed point. In four dimensions, the signature of holographic theories possessing general relativity duals is that the $a$ and $c$ anomaly coefficients are equated (at large $N$). This is reproduced through the aforementioned Hamiltonian methods in the fifth section. 
\section{Comparison to earlier work}
The following work is an amalgamation of several related ideas and in order to point out what the novelty of the results are, it will help to put it in context with earlier work. 

I will start with the idea of mapping renormalization group flows to Hamiltonian systems. This was first done by Dolan in \cite{Dolan} where a dictionary to map renormalised beta function and Callan--Symanzik equations to a Hamiltonian system was presented. In the case of large $N$ matrix field theories, efforts to rewrite the RG flow equations as Hamilton--Jacobi equations and to identify a Hamiltonian from the exact RG equations were made in the work of Becchi et. al. in \cite{Imbimbo2}, Akhmedov in \cite{Akhmedov} and references therein. All the works dealt with the Wilsonian RG flow of regulated yet un-renormalized quantities, but a dictionary very similar to the one laid out by Dolan seems to still apply, and this point is elucidated upon in \cite{Papa2}. 

The same kind of mapping can be seen to result from the semiclassical limit of the quantum RG introduced by Sung--Sik Lee in \cite{S.S.Lee1},\cite{S.S.Lee2}. This is a more general procedure that aims to map the RG flow of large $N$ matrix field theories into quantum field theories in one higher dimension where a subset of the scale dependent sources play the role of dynamical fields. The author's addendum to this body of work was to recognise the key role the emergence of diffeomorphism invariance in the dual bulk theory plays in defining Holographic theories \cite{me} (more on this in what is to follow).

The formulation of holographic renormalization in the language of canonical transformations first appeared in \cite{Papa1}. This method is dubbed the Hamiltonian approach to holographic RG.  The systematic algorithm to solve the
Hamilton-Jacobi equation based on the dilatation operator was put
forward in \cite{PapaSken}, and to see a review of both these results, see \cite{Papa2}. The results of standard Holographic RG as formulated in \cite{HRG} for example are reproduced including the computation of the Holographic anomaly (as was first done in \cite{Skenderis}). All this features in section 5. 

The novelty of the results presented in this article is to start with deducing the off shell bulk theory of general relativity coupled to matter by combine the insights of mapping RG flows of large $N$ theories to Hamiltonian systems and also that the emergence of general covariance as encoded in the so called Holographic Wess--Zumino consistency conditions is a defining feature of holographic dualities. This approach sidesteps the path integral construction in the quantum RG as it applies mainly to the large $N$ limit. Then, the Hamiltonian approach to holographic RG can be applied in order to study the theory at UV fixed point leading to identical results of conventional Holographic renormalization in the AdS/CFT context. 

\section{Mapping RG flows to Hamiltonian Systems}

The renormalization group flow equations are a statement of the independence of physical observables in quantum field theories under change of scale. As is customary, observables such as correlation functions are computed through the intermediary of the partition function $Z[J^{\alpha}(x),g_{\mu\nu}(x)]$ where the coupling constants are upgraded into space dependent sources $J^{\alpha}(x)$ including that of the composite operators like the energy momentum tensor which couples to $g_{\mu\nu}(x)$. Correlations functions are computed through taking functional derivatives of this partition function with respect to the appropriate sources. For now, these sources and operators are un-renormalized, although it is assumed that the partition function is appropriately regulated. The following results can be arrived at through the semiclassical limit of a procedure that constructs a path integral for the quantum theory in the bulk known as the quantum renormalization group \cite{S.S.Lee1}, \cite{S.S.Lee2}. 

\subsection{The Phase Space}
For renormalization group flows to be related to some Hamiltonian system, an identification of the phase space variables need be made. One obvious candidate is the source $J(x)$, and the conjugate is the one point function in this source's presence:
\begin{equation}\langle\mathcal{O_{\alpha}}(x)\rangle_{J}=\frac{\delta \textrm{ln}Z}{\delta J^{\alpha}(x)}.\label{orig}\end{equation}
And so the symplectic form on this phase space reads 
\begin{equation}\Omega=\int\textrm{d}^{D}x \sqrt{g} \delta \langle \mathcal{O}_{\alpha}(x)\rangle \wedge \delta J^{\alpha}(x).
 \end{equation}
 Then, it isn't hard to see that the generating functional $W[J^{\alpha}(x),g_{\mu\nu}(x)]=\textrm{ln}Z[J^{\alpha}(x),g_{\mu\nu}(x)]$ plays the role of Hamilton's principal function, or the on shell action. Hence, the symplectic form evaluated on shell vanishes:
 $$\Omega|_{\langle\mathcal{O}\rangle=\frac{\delta W}{\delta J}}=0.$$
For notational convenience, henceforth I introduce the notation, $P_{\alpha}(x)\equiv \langle \mathcal{O}_{\alpha}(x)\rangle$.
In all that follows, the index $\alpha$ runs over all operators and sources, so it is a multi index set as it encomapsses sources and operators potentially with different numbers of spacetime indices. 
 In the case of the renormalised theory, these identifications were first made by Dolan in \cite{Dolan}. This identification for the regularised theory was also made in \cite{Papa2}. 

More shall be said about this in a later section where the Hamilton--Jacobi theory is described in detail. Now I wish to identify the Hamiltonian which generates evolution in scale. 

\subsection{The Hamiltonian}
In order to deduce which function on the previously identified phase space drives the RG flow, one can exploit the fact that the generating functional plays the role of Hamilton's principal functional. This is analogous to the treatment of this matter in \cite{Papa2}. Then the Hamilton--Jacobi equation reads:
\begin{equation}
H[J,P]=\frac{\partial W[J]}{\partial \tau}. 
\end{equation}
Recalling that Hamilton's principal function is nothing but the on shell action, a Lagrangian too can be identified as
\begin{equation}
L[J,P]=\frac{\textrm{d}W[J]}{\textrm{d}\tau}=\int \textrm{d}^{D}x\sqrt{g} \left(P_{\alpha}\dot{J}^{\alpha}-H[J,P]\right).
\end{equation}
Hamilton's equations then read:
\begin{equation*}
\beta^{\alpha}(J)\equiv\dot{J}^{\alpha}(x)=\frac{\delta H}{\delta P_{\alpha}}
\end{equation*}
\begin{equation*}
\dot{P}_{\alpha}\equiv \frac{\partial \langle \mathcal{O}_{\alpha}\rangle }{\partial \tau}=-\frac{\delta H}{\delta J^{\alpha}}
\end{equation*}
When dealing with an unrenormalized generating functional, which shall be the object of considerations in section 4, we note that its scale dependence is carried solely by the sources it depends on. This implies that there is no explicit dependence on the flow time in the generating functional, and conseqently 
\begin{equation*}
H[P,J]=\frac{\partial W[J]}{\partial \tau}=0.
\end{equation*}
In other words, the Hamiltonian is totally constrained. Given the beta functions, integrating the first Hamilton's equation yields a more explicit form:
\begin{equation}
H[P,J]=\int \textrm{d}^{D}x\sqrt{g} \sigma(\tau) \left(\beta^{\alpha}(J)P_{\alpha}+U(J)\right)=0. \label{constlapse}
\end{equation} 

The function $\sigma(\tau)$ inserted into this expression is the lagrange multiplier enforcing the Hamiltonain constraint. The invariance this constraint generates is an RG version of re-parameterization invariance. This is invariance under the re-labelling $\tau\rightarrow f(\tau)$, and the role of $\sigma(\tau)$ is to ensure that the analogue of `proper time' $\int\sigma(\tau)\textrm{d}\tau$ remains invariant under such changes. The function $U(J)$ is related to the beta function encoding the renormalization of the identity operator. In order to see this, note the explicit version of the second Hamilton's equation:
\begin{equation}
-\frac{\partial P_{\alpha}}{\partial \tau}=\frac{\partial \beta^{\kappa}(J)}{\partial J^{\alpha}}P_{\kappa}+\frac{\partial U}{\partial J^{\alpha}}.
\end{equation}
Recalling that $P_{\alpha}=\langle \mathcal{O}_{\alpha}\rangle$, we see that this equation is the Callan--Symanzik equation describing the scale evolution of the one point function in the presence of sources $J^{\alpha}$. The derivative of the beta function near a fixed point defines the matrix of anomalous dimensions and in analogy, the second term too can be seen as encoding the renormalization of the coupling of the identity operator with $U(J)$ playing the role of the beta function. 

When dealing with the generating functional of a renormalized theory, the condition of relevance is that it doesn't depend on the scale at all:
\begin{equation*}
\frac{\textrm{d}}{\textrm{d}\tau} W^{\textrm{ren}}[J]=0.
\end{equation*}
This implies the vanishing of the Lagrangian:
\begin{equation}
L=0= \left(\int \textrm{d}^{D}x\sqrt{g}\beta^{\alpha}(J)P^{\textrm{ren}}_{\alpha}\right)-\frac{\partial }{\partial \tau}W^{\textrm{ren}}[J].
\end{equation}
The explicit dependence of the generating functional on the RG scale is carried by the anomaly:
\begin{equation}
\frac{\partial}{\partial \tau}W^{\textrm{ren}}[J]=\mathcal{A}[J].
\end{equation}
The symbol $\mathcal{A}[J]$ denotes the integrated anomaly: $\mathcal{A}=\int \textrm{d}^{D}x \sqrt{g} A(J)$, where $A(J)$ is a local functional of the sources and their spatial derivatives. 
This then defines a true Hamiltonian:
\begin{equation}
H_{\textrm{ren}}[P^{\textrm{ren}},J]=\int \textrm{d}^{D}x\sqrt{g} \sigma(\tau)\beta^{\alpha}(J)P^{\textrm{ren}}_{\alpha}=-\mathcal{A}_{\sigma}[J].\label{renham}
\end{equation}
The object on the right hand side of the second equality is defined as the integrated anomaly smeared against $\sigma$, i.e. $\mathcal{A}_{\sigma}[J]=\int \textrm{d}^{D}x \sigma A(J)$. 

The step of renormalization involves going from the phase space with canonical variables $(J^{\alpha}(x),P_{\alpha})$ and Hamiltonian $H[P,J]$ to a phase space with variables $(J^{\alpha},P^{\textrm{ren}}_{\alpha})$ and Hamiltonian $H_{\textrm{ren}}[P^{\textrm{ren}},J]$ and is accomplished via a canonical transformation as noticed in \cite{Papa1}. 
\subsection{Local Renormalization Group}
The most general notion of coarse graining available in real space which remains meaningful even on arbitrary backgrounds is the one given by local Weyl transformations of the background metric and of the other sources. The renromalization group flow can be seen as the response of the generating functional under such a change. This perspective is known as the local renormalization group. It is a continuum generalisation of Kadanoff's idea of block spin transformations. This approach was pioneered by Osborn in \cite{Osborn1}. 

This generalization can be summarised by allowing the function $\sigma$ now to also depend on space. First, note that
under general scale transformations, the set of sources $\left\{J^{\hat{\alpha}}\right\}$ that includes the sources for the energy momentum tensor and the identity operator are transformed in a manner involving beta functions: $J^{\hat{\alpha}}(x)\rightarrow \Delta_{\sigma}J^{
\hat{\alpha}}=-\sigma(x,\tau)\beta^{\hat{\alpha}}(J(x,\tau))$.
This response is such that the generating functional invariant:
\begin{equation}
\Delta_{\sigma}W[J]=\int \textrm{d}^{D}x \sqrt{g} \sigma(x,\tau)\beta^{\hat{\alpha}}(J)\frac{\delta W[J]}{\delta J^{\hat{\alpha}}}=0. \label{lrgg}
\end{equation}

Note that here too the generating functional remains unrenormalized. Given that the identity operator too is included now in the set of sources, and that the function $U(J)$ encodes the flow of the coupling of the identity operator, the Hamiltonian now reads
\begin{equation}
H(\sigma;P,J]=\int \textrm{d}^{D}x \sqrt{g}\sigma(x,\tau)\beta^{\hat{\alpha}}(J)P_{\hat{\alpha}}(x)=0.
\end{equation}
The fact that this is a constraint, meaning it vanishes, reflects the condition \eqref{lrgg}. \footnote{To see why this is just the same as the unrenormalized Hamiltonian \eqref{constlapse}, one need only separate the identity sectors' contribution to the above expression:
\begin{equation*}
H(\sigma;P,J]=\int \textrm{d}^{D}x \sqrt{g}\sigma(x,\tau)\left(\beta^{\hat{\alpha}'}(J)P_{\hat{\alpha}'}(x)-U(J)\right),
\end{equation*}
where the primed indices don't include the identity operator.}
This will be important in the later sections. For notational ease, I will drop the dependence of the Hamiltonian on the phase space variables and write it as $H(\sigma)$ henceforth. 
Notice that the above expression implies that the metric too has a beta function term, i.e. $$\Delta_{\sigma}g_{\mu\nu}=-\sigma(x)\beta_{\mu\nu}(g),$$
which seems somewhat odd from the conventional framing of the local RG. This is an artefact of the fact that this equation deals with the unrenormalized generating functional. When dealing with a renormalized theory in the vicinity of the fixed point, this equation reads
\begin{equation}
\Delta_{\sigma}W^{\textrm{ren}}[J]=\int \textrm{d}^{D}x \sqrt{g} \sigma(x,\tau)\left(-g_{\mu\nu}\frac{\delta }{\delta g_{\mu\nu}}+\beta^{\hat{\alpha}'}(J)\frac{\delta }{\delta J^{\hat{\alpha}'}}\right)W^{\textrm{ren}}[g,J]=\mathcal{A}_{\sigma}[g,J].
\end{equation} 
The primed index set excludes the metric. This is the form of the local renormalization group equation most often encountered in the literature. Notice that here, the metric beta function if expanded in derivatives to orders two or higher, i.e. anything beyond $\beta_{\mu\nu}\propto g_{\mu\nu}$ doesn't feature. 
Given this setup, I will proceed to apply it to field theories possessing gravity duals in the bulk. 

\subsection{Large $N$ Matrix Field Theories}
Consider a theory with $N$ degrees of freedom (where $N$ could for example be the rank of the gauge group) organised as matrix fields. Also, let us pay heed to the case where $N$ is taken to infinity. The advantages this situation has to offer is that there is an effective factorisation of one point functions of a certain class of singlet composite operators known as `multi trace' operators into a class of operators known as `single trace' operators. Beyond the large $N$ limit there is but suppression of these multi trace operators by powers of $1/N$. 

To illustrate this, I will consider the one point function of such a generic multi trace operator at large $N$. This factories as
\begin{equation}
\langle \mathcal{O}_{\tilde{\alpha}} \rangle|_{N\rightarrow \infty}= \mathlarger{\mathlarger{\sum}}_{ \stackrel{\mathlarger{(\hat{\alpha})} }{\ \, (\mu),(\nu),(\rho),\cdots}}F\big(j\big)_{(\mu),(\nu),(\rho),\cdots,}^{\left\{(\hat{\alpha}_{i}),(\hat{\alpha}_{j}),(\hat{\alpha}_{k}),\cdots\right\}}\nabla^{(\mu)}\langle O_{\hat{\alpha}_{i_{1}}\hat{\alpha}_{i_{2}}\cdots}\rangle\nabla^{(\nu)}\langle O_{\hat{\alpha}_{j_{1}}\hat{\alpha}_{j_{2}}\cdots}\rangle\nabla^{(\rho)}\langle O_{\hat{\alpha}_{k_{1}}\hat{\alpha}_{k_{2}}\cdots}\rangle\cdots
\end{equation}
$\left\{\mathcal{O}_{\tilde{\alpha}}\right\}$ denotes the set of all composite singlet operators, single and multi-trace except the identity.
The multi index notation works as follows: The symbol $\nabla_{(\mu)}$ is short hand for $\nabla_{\mu_{1}}\nabla_{\mu_{2}}\nabla_{\mu_{3}}\cdots$. The sets $\left\{O_{(\hat{\alpha})}\right\}$ refers to the single trace operators, and the multi indices in parantheses $(\hat{\alpha}_{i})$ idenote the sets $\left\{\hat{\alpha}_{i_{1}}\hat{\alpha}_{i_{2}}\cdots,\hat{\alpha}_{j_{1}}\hat{\alpha}_{j_{2}}\cdots,\hat{\alpha}_{k_{1}}\hat{\alpha}_{ik{2}}\cdots, \cdots\right\}$. 

This factorisation property implies something very interesting when the RG Hamiltonian is considered. The phase space variables in the large $N$ limit will go from the set of sources of all singlet operators and the expectation values of such operators down to the sources and expectation values of single trace operators only. The cost of this reduction is that the Hamiltonian no longer takes a simple form. To see this, consider the RG Hamiltonian for such theories in the large $N$ limit:
$$H(\sigma)|_{N\rightarrow \infty}=\int \textrm{d}^{D}x \sqrt{g} \sigma(x) \left(\beta^{\tilde{\alpha}}(J)P_{\tilde{\alpha}}-U(J)\right)|_{N\rightarrow \infty}$$
\begin{equation}=\int \textrm{d}^{D}x \sqrt{g} \sigma(x) \left(\mathcal{F}(p_{(\hat{\alpha})},j^{(\hat{\alpha})})-U(j)\right).\end{equation}
The identity operator's contribution has been separated and thus the multi indices do not include it. The canonical variables now defined by the sources and expectation values of the single trace operators only are denoted $(p_{(\hat{\alpha})},j^{(\hat{\alpha})})$.

The function $\mathcal{F}(p_{(\hat{\alpha})},j^{(\hat{\alpha})})$ results from the expansion of the expectation values of the general, multi trace operators into single trace operators:
\begin{equation} P_{\tilde{\alpha}}|_{N\rightarrow \infty}= \mathlarger{\mathlarger{\sum}}_{ \stackrel{\mathlarger{(\hat{\alpha})} }{\ \, (\mu),(\nu),(\rho),\cdots}}F\big(j\big)_{(\mu),(\nu),(\rho),\cdots}^{\left\{(\hat{\alpha}_{i}),(\hat{\alpha}_{j}),(\hat{\alpha}_{k}),\cdots\right\}}\nabla^{(\mu)} p_{\hat{\alpha}_{i_{1}}\hat{\alpha}_{i_{2}}\cdots}\nabla^{(\nu)} p_{\hat{\alpha}_{j_{1}}\hat{\alpha}_{j_{2}}\cdots}\nabla^{(\rho)} p_{\hat{\alpha}_{k_{1}}\hat{\alpha}_{k_{2}}\cdots}\cdots\end{equation}
The various momentum independent functions $F(j)_{(\mu),(\nu),(\rho),\cdots,}^{\left\{(\hat{\alpha}_{i}),(\hat{\alpha}_{j}),(\hat{\alpha}_{k}),\cdots\right\}}$ can be thought of as the beta functions for various higher trace operators. Similarly the potential $U(j)$ depends on just the single trace sources $j^{(\hat{\alpha})}(x)$ and its derivatives. 

Thus, the Hamiltonian in the large $N$ limit effectively describes the projected RG flow down to the subspace of single trace operators. However, the dynamics it generates is much more non trivial than for generic RG flows in that it is no longer linear in the momentum. This means that if one can find conditions under which this Hamiltonian is restricted to be quadratic in the momenta say, then the RG flow can be seen as describing the dynamical evolution of a system with second order equations of motion in configuration space. The system it describes also seems to inhabit one higher dimension than the quantum field theory whose RG flow is being studied. These are all the features one expects to find for holographic RG flows of quantum field theories possessing bulk duals. Naturally, the scale dependence of the sources plays the role of the extra `radial' direction as one would expect. 

Such an identification of a Hamiltonian for renormalization group flows of matrix field theories was also made in \cite{Imbimbo2}, \cite{Akhmedov}. 
\section{Holographic RG flows}
Large $N$ matrix field theories whose RG flow can be mapped into a dynamical Hamiltonian which is quadratic in momenta through the aforementioned procedure will be of primary interest. This is the case for holographic theories in regimes where the bulk dynamics is that of general relativity coupled to matter. In addition to possessing quadratic Hamiltonians, such theories are also assumed to be those for which one is allowed to truncate the infinite set of single trace operators to a set containing only a finite number of them. The mechanism behind such a truncation is a gap in the spectrum of anomalous dimensions which scales with $N$ in such a way that in the large $N$ limit the only operators with finite scaling dimensions are those in the finite set mentioned above. 
I will show for a simple setting that the quadratic nature of the Hamiltonian follows form an additional consistency condition for the local renormalization group, which I call the Holographic Wess--Zumino consistency conditions. 
\subsection{Pure gravity in the bulk}

For simplicity, I will take this set to contain just one operator, which is the single trace energy momentum tensor of the matrix field theory $T_{\mu\nu}(x,\tau)$. The dual theory, living in one higher dimension shall be one of a dynamical metric $g_{\mu\nu}(x,\tau)$ (with conjugate momentum $\pi^{\mu\nu}(x,\tau)\equiv\langle T^{\mu\nu}(x,\tau)\rangle_{g}$). 
The phase space for the bulk theory is thus the one spanned by the variables $(g_{\mu\nu},\pi^{\mu\nu})$ and tentatively, the Hamiltonian must take the form 
$$H(\sigma)=\int \textrm{d}^{D}x \sqrt{g}\sigma(x,\tau)\left(\mathcal{F}(\pi,g)+U(g)\right).$$

There is more than just the RG Hamiltonian, which one can think of as the anomalous Ward identity corresponding to violated Weyl invariance, in the case of this theory. The diffeomorphism Ward Identity too must be translated into a constraint which is added to the total Hamiltonian. The fact that constraints in the bulk are related to the Ward identities of the quantum field theory was noticed by Corley in \cite{Corley}. The connection between the local renormalization group and the holographic renormalization group was also discussed in \cite{Erd}. 
The diffeomorphism Ward identity is expressed as the covariant conservation of the expectation value of the Energy momentum tensor, which on this phase space reads
$$\nabla_{\mu}\langle T^{\mu\nu}\rangle_{g}=\nabla_{\mu}\pi^{\mu\nu}=0,$$
and so this can be written as a vector constraint
\begin{equation}H_{\mu}(\xi^{\mu})=-2\int \textrm{d}^{D}x \sqrt{g} \xi^{\mu}\nabla_{\nu}\pi^{\nu}_{\mu}.\label{vc}\end{equation}
Through its Poisson algebra it does represent the algebra of diffeomorphisms as one would expect:
\begin{equation}\left\{H_{\mu}(\xi^{\mu}),H_{\nu}(\zeta^{\nu})\right\}=H_{\mu}([\xi,\zeta]^{\mu}). \label{hc1}\end{equation}

The demand that the scalar Hamiltonian constraint respect this invariance is ensured thought the fact that the Hamiltonian density is a tensor density of weight one. For any such density, the Poisson brackets with the above vector constrained are fixed, and in particular, we have
\begin{equation} \left\{H(\sigma),H_{\mu}(\xi^{\mu})\right\}=H(\xi^{\mu}\partial_{\mu}\sigma).\label{hc2}\end{equation}

In order to further fix the form of the functions $\mathcal{F}(\pi,g)$ and $U(g)$, which would then fix the form of the Hamiltonian, we need consider the Poisson bracket between two scalar constraints. Here is where positing that diffeomorphism invariance currently manifest on the constant radius hypersurfaces need be upgraded into diffeomorphism invariance in the full $D+1$ dimensional bulk. This translates into a very specific manner in which the Poisson brackets between two scalar constraints needs to close:
\begin{equation}\left\{H(\sigma),H(\sigma')\right\}=H_{\mu}(g^{\mu\nu}(\sigma\partial_{\nu}\sigma'-\sigma'\partial_{\nu}\sigma)).\label{pb}\end{equation}
These Poisson bracket relations reflect the Lie bracket structure of spacetime vector fields that are decomposed tangentially and orthogonally to the hypersurface. These components describe the normal and tangential deformations of hypersurfaces embedded in spacetime as is discussed in detail in \cite{Teitelboim}.

The Poisson bracket relation \eqref{pb} which translated back into field theory terms becomes a particular manner in which the Wess--Zumino consistency conditions of the local renormalization group need be satisfied:
\begin{equation}0=[\Delta_{\sigma},\Delta_{\sigma'}]\textrm{ln}Z[g]|_{N\rightarrow \infty}=\int \textrm{d}^{D}x g^{\mu\nu}(\sigma\partial_{\nu}\sigma'-\sigma'\partial_{\nu}\sigma) \nabla^{\sigma}\frac{\delta \textrm{ln}Z[g]}{\delta g^{\mu\sigma}}.\end{equation}
This is what I call the Holographic Wess--Zumino consistency condition. This was first posited in \cite{me}. These considtions are the field theoretic codification of bulk diffeomorphism invariance. 
These Poisson bracket relations restrict the functional forms of $\mathcal{F}(\pi,g)$ and $U(g)$. There are two cases of interest:

\subsubsection{General Relativity}
When the number of dimensions $D\leq 4$, there is in fact a unique function on phase space up to canoncial transformations (see next subsection) that realizes this Poisson bracket relation, which is of the form:
\begin{equation}H(\sigma)=\int \textrm{d}^{D}x  \sigma(x)\left(\frac{1}{\sqrt{g}}\left(\pi_{\mu\nu}\pi^{\mu\nu}-\frac{1}{D-1}\textrm{tr}\pi^{2}\right)-\sqrt{g}(R-2\Lambda)\right).\label{hamc}\end{equation}
This is nothing but the scalar Hamiltonian constraint of general relativity first written down in \cite{ADM}. Thus the demand for emergent diffeomorphism invariance fixes the off shell dynamics of the bulk theory to be that which GR describes. The theorem that ensures this is that of Hojman, Kuchar and Teitelboim in \cite{HKT}. For notational convenience it is worth noting that the term quadratic in the momenta can be written more compactly by introducing the de- Witt supermetric:
\begin{equation*}
G_{\mu\nu\rho\sigma}\equiv g_{\mu\rho}g_{\nu\sigma}+g_{\mu\sigma}g_{\nu\rho}-\frac{1}{D-1}g_{\mu\nu}g_{\rho\sigma},
\end{equation*}
and so
\begin{equation*}
\pi^{\mu\nu}\pi_{\mu\nu}-\frac{1}{D-1}\textrm{tr}\pi^{2}=G_{\mu\nu\rho\sigma}\pi^{\mu\nu}\pi^{\rho\sigma}.
\end{equation*}
Without going into details of the proof that this is the unique function that satisfies the constraint algebra, the idea is as follows: the Poisson brackets maintain the number of derivatives while decreasing the total power of the momenta by one. Also, the only non vanishing terms are those where derivatives hit the smearing functions so terms ultralocal in the metric and the momenta commute. To see how these facts might aid with proving this result, consider the bracket:
\begin{equation*}
\left\{H(\sigma),H(\sigma')\right\}=\left\{\int_{x} \sigma(x)(\mathcal{F}(g,\pi)+U(g)),\int_{y}\sigma'(y)(\mathcal{F}(g,\pi)+U(g))\right\}
\end{equation*}
$$=\int \textrm{d}^{D}x \int \textrm{d}^{D}y  \bigg(\left\{\sigma(x)\mathcal{F}(g(x),\pi(x)),\sigma'(y)\mathcal{F}(g(y),\pi(y))\right\}+$$ 
$$\left\{ \sigma(x)\mathcal{F}(g(x),\pi(x)),\sigma'(y)U(g(y))\right\}-(\sigma\leftrightarrow \sigma')\bigg).$$
If the function $\mathcal{F}(g,\pi)$ contains $n$ momenta, then the first terms yields a term with $2n-1$ momenta and the second yeilds a term with $n-1$ momenta.
The right hand side reads $$\int \textrm{d}^{D}x \sqrt{g} g^{\mu\nu}(\sigma\partial_{\nu}\sigma'-\sigma'\partial_{\nu}\sigma)\nabla_{\kappa}\pi^{\kappa}_{\mu}.$$
It contains two derivatives and one power of the momentum. The most minimal prescription for this to be satisfied would be that the $n=1$ and $\mathcal{F}(g,\pi)$ contains one derivative, except there is no way to form a scalar from just these variables. The other alternative is that there are no gradients in the function and it is thus ultralocal. Then, looking at the second line involving the bracket between $\mathcal{F}$ and $U$, one sees that this must be a quadratic ultralocal function at most. 

Then, working through the bracket and comparing it to $H_{\mu}(g^{\mu\nu}(\sigma\partial_{\nu}\sigma'-\sigma'\partial_{\nu}\sigma))$ whose form is known, one finds that 
\begin{equation}
\mathcal{F}(g,\pi)=\frac{1}{\sqrt{g}}\left(\pi_{\mu\nu}\pi^{\mu\nu}-\frac{1}{D-1}\textrm{tr}\pi^{2}\right),
\end{equation}
and 
\begin{equation}
U(g)=-\sqrt{g}(R-2\Lambda).
\end{equation}
Note that the reltative sign between the `kinetic' term $\mathcal{F}(g,\pi)$ and the `potential' $\sqrt{g}(R-2\Lambda)$ originating from the sign on the right hand side of \eqref{pb} reflects the spacelike nature of the radial direction.
\subsubsection{Lovelock type higher derivative theories}
In dimensions 4 and higher, in all generality, theories of a dynamical metric posessing second order equations of motion in configuration space and respecting diffeomorphism invariance are of the Lovelock type \footnote{In this section, I follow the treatement of \cite{Liu} and \cite{BunstZan}}. This means that in the bulk, in addition to the Einstein Hilbert term, there are very specific higher curvature terms of the following form:
\begin{equation}
S=\int \textrm{d}^{D+1}x \sqrt{\gamma}(\Lambda+\,^{(D+1)}R(\gamma)+\sum^{\frac{D+1}{2}}_{k=1}S_{k})\label{sll}
\end{equation}
where 
\begin{equation*}
S_{k}=-\frac{1}{2k!}\int \textrm{d}^{D+1}x \sqrt{\gamma}\delta^{a_{1}\cdots a_{2k}}_{b_{1}\cdots b_{2k}}\,^{(D+1)}R^{b_{1}b_{2}}_{a_{1}a_{2}}\cdots \,^{(D+1)}R^{b_{2k-1}b_{2k}}_{b_{2k-1}b_{2k}}
\end{equation*}
The indices $a,b,\cdots$ run from $0$ to $D+1$ and bulk spacetime metric is denoted $\gamma_{ab}$. The superscript $\,^{(D+1)}$ indicates that the curvature tensors are of the bulk spacetime. 

Each term in the sum is an Euler characteristic in dimensions lower than $D+1$. The Hamiltonian analysis of these theories is of interest here. In order to see the difficulties this presents, it is worth limiting our attention to the $D=4$ case, which will also be of relevance to the next section. 
In 5 bulk dimensions the higher curvature contribution to the action is the Gauss-Bonnet invariant:
\begin{equation*}
S=\int \textrm{d}^{D+1}x \sqrt{\gamma}\left(\Lambda+\,^{(D+1)}R+\alpha\left(\,^{(D+1)}R^{2}-4\,^{(D+1)}R_{ab}\,^{(D+1)}R^{ab}+\,^{(D+1)}R_{abcd}\,^{(D+1)}R^{abcd}\right)\right).
\end{equation*}
Now, in order to study the Hamiltonian formulation, it is necesary to perform a $(D+1)$ split of the above action and then perform the Legendre transform. 

The spacetime metric is related to the metric on a $r=const$ hypersurface with normal $n_{a}$ through the relation $\gamma_{ab}=n_{a}n_{b}-g_{ab}$ and the $D$ dimensional diffeomorphism covariant version of the `velocity' field for the hypersruface metric is given by the extrinsic curvature tensor: $K_{ab}=\nabla_{(a}n_{b)}$ its components lie entirely tangential to the hypersruface and it can alternatively be written as $K_{\mu\nu}=\mathcal{L}_{n}g_{\mu\nu}$, i.e. the Lie derivative of the metric on the hypersurface with respect to the normal. The key identity is the one relating the hypersurface tangential components of the bulk curvature tensor to the intrinsic and extrinsic curvature tensors on the hypersurface 
\begin{equation}
\,^{(D+1)}R_{\mu\nu\rho\sigma}= R_{\mu\nu\rho\sigma}+K_{\mu\sigma}K_{\nu\rho}-K_{\mu\rho}K_{\nu\sigma}
\end{equation}
 In terms of $(K_{\mu\nu},g_{\mu\nu})$, the above action is written as
\begin{equation*}
S=\int\textrm{d}^{D+1}x \sqrt{\gamma}\bigg(\Lambda+R+K^{2}-K_{\mu\nu}K^{\mu\nu}+\alpha\bigg((R+K^{2}-K_{\mu\nu}K^{\mu\nu})^{2}-4(R_{\mu\nu}+KK_{\mu\nu}-K_{\mu\kappa}K^{\kappa}_{\nu})^{2}\end{equation*} $$+(R_{\mu\nu\rho\sigma}+K_{\mu\sigma}K_{\nu\rho}-K_{\mu\rho}K_{\nu\sigma})^{2}-\frac{4}{3}K^{4}+8K^{2}K_{\mu\nu}K^{\mu\nu}-\frac{32}{3}KK^{\mu}_{\sigma}K^{\rho}_{\mu}K^{\sigma}_{\rho}-4(K_{\alpha\beta}K^{\alpha\beta})^{2}+8K^{\mu}_{\nu}K^{\nu}_{\sigma}K^{\sigma}_{\rho}K^{\rho}_{\mu}\bigg),$$
it is easy to see that in the $\alpha\rightarrow0$ limit, one recovers the action of General relativity involving only the Ricci scalar and the cosmological constant. Now, in order to performa Legendre transform, one first identifies the Momentum conjugate to the metric on the hypersurface defined by
\begin{equation*}
\pi_{\mu\nu}=\frac{\partial L}{\partial K_{\mu\nu}},
\end{equation*}
where the Lagrangian is identified as the integrand of the action: $S=\int \textrm{d}^{D+1}x L(g,K)$. 
This momentum can be readily computed explicitly:
\begin{equation*}
\pi_{\mu\nu}=K_{\mu\nu}-Kg_{\mu\nu}-2\alpha\bigg(g_{\mu\nu}(RK-2R_{\mu\nu}K^{\mu\nu})-RK_{\mu\nu}-2R_{\mu\nu}K+4R^{\gamma}_{(\mu}K_{\nu)\gamma}+ \end{equation*} 
$$+ 2R_{\mu\nu\rho\sigma}K^{\mu\rho}+\frac{1}{3}g_{\mu\nu}(-K^{3}+3KK_{\alpha\beta}K^{\alpha\beta}+2K^{\delta}_{\gamma}K^{\eta}_{\delta}K^{\gamma}_{\eta})+K^{2}K_{\mu\nu}-$$ \begin{equation}2KK^{\gamma}_{\mu}K_{\gamma\nu}-K_{\mu\nu}K_{\alpha\beta}K^{\alpha\beta}+2K^{\sigma}_{\mu}K^{\rho}_{\sigma}K_{\rho\nu}\bigg), \label{moK}\end{equation}
and the Hamiltonian density is then defined as
\begin{equation*}
H(g_{\mu\nu},\pi^{\mu\nu})=\pi^{\mu\nu}K_{\mu\nu}-L(K_{\mu\nu},g_{\mu\nu}),
\end{equation*}
and it is understood that the relationship between the momentum and the extrinsic curvature need be inverted in order to determine this function. There is however an obstacle to doing so, which is readily seen by noticing that the right hand side of \eqref{moK} is cubic in the extrinsic curvature tensors. This implies that if somehow this equation can be solved for $\pi_{\mu\nu}$, there could be one or three distinct values of $K_{\mu\nu}$ corresponding to a given value of $\pi_{\mu\nu}$. This leads to a so called `branched' or multi-valued Hamiltonian which could also possess cusps. This problem was noted in \cite{DF} and \cite{BunstZan} originally. 

It is however possible to invert the relationship between the momentum and extrinsic curvature if the higher order terms are treated perturbatively with $\alpha$ playing the role of the necessary small parameter. Then a Hamiltonian can be found to $O(\alpha^{2})$. Analysis along these lines was performed in \cite{Liu} with the goal of computing counterterms in an asymptotically AdS space. Consequently the constraint algebra is also satisfied up to corrections of some order in $\alpha$ for such a Hamiltonian and not exactly. Thus such an RG Hamiltonian doesn't strictly satisfy the consistency conditions. 

A more robust method to deal with this issue is to treat the Gauss--Bonnet theory as one would a more general higher derivative theory. This involves treating the configuration variable, i.e. the metric and it's velocity, here the extrinsic curvature as independent configuration like variables and finding momenta for both these variables. Then for theories such as Gauss--Bonnet gravity that ought to possess the same number of degrees of freedom as general relativity, there are additional constraints on this enlarged phase space that eliminte the right number of auxiliary degrees of freedom for this to be accomplished. A resolution of the branched nature of the Hamiltonian along these lines but in addition to also adding other non Gauss--Bonnet type higher derivative terms, was described in \cite{Branch} in the minisuperspace context. 

In fact the introduction of new degrees of freedom allows one to treat a variety of higher derivative theories of gravity in the Hamiltonian formalism, and there too if the corrections beyond general relativity are treated perturbatively, one may obtain an approximate `pseudo Hamiltonian' on the phase space of general relativity. This procedure is described in detail in \cite{Jp}. There too however, it was noted that such a Hamiltonian is not the true Hamiltonian of the system as it generates different evolution from the full Hamiltonian living on the extended phase space. 

\subsubsection{Summary}
Starting with noticing that the Poisson bracket relations:
\begin{eqnarray*}
\left\{H_{\mu}(\xi^{\mu}),H_{\nu}(\zeta^{\nu})\right\}=H_{\mu}([\xi,\zeta]^{\mu}),\\
\left\{H(\sigma),H_{\mu}(\xi^{\mu})\right\}=H(\xi^{\mu}\partial_{\mu}\sigma),\\ \,\,\left\{H(\sigma),H(\sigma')\right\}=H_{\mu}(g^{\mu\nu}(\sigma\partial_{\nu}\sigma'-\sigma'\partial_{\nu}\sigma)),
\end{eqnarray*}
represent the manifestation of spacetime diffeomorphism invariance on the phase space with canonical co-ordiantes $(g_{\mu\nu},\pi^{\mu\nu})$. For $D\leq 3$, the unique set of phase space fucntions $H(\sigma), H_{\mu}(\xi^{\mu})$ that satisfy this algebra are those of general relativity given by
\begin{eqnarray*}
H(\sigma)=\int \textrm{d}^{D}x  \sigma(x)\left(\frac{1}{\sqrt{g}}\left(\pi_{\mu\nu}\pi^{\mu\nu}-\frac{1}{D-1}\textrm{tr}\pi^{2}\right)-\sqrt{g}(R-2\Lambda)\right),\\
H_{\mu}(\xi^{\mu})=-2\int \textrm{d}^{D}x \sqrt{g} \xi^{\mu}\nabla_{\nu}\pi^{\nu}_{\mu},
\end{eqnarray*}
which are the constraint of general relativity in the ADM formalism. In higher dimensions however, the most general diffeomorphism invariant theories posessing second order equations of motion in the Lagrangian formulation are the Lovelock type theories with action given by \eqref{sll}. However, on the phase space co-ordinatised by $(g_{\mu\nu},\pi^{\mu\nu})$, the dynamics generated by the Hamiltonian obtained from the Legendre transform of the Lagrangian in \eqref{sll} is problematic. Specifically, this Hamiltonian is a branched or multivalued fucntion of the momenta with cusps. The dynamics generated by such Hamiltonians is generically pathological as the flow generated by such a Hamiltonian could jump from one branch to the other in the course of the evolution it generates. Potential resolutions of this difficulty necessarily invovles enlarging phase space through the inclusion of new fields. This is despite the fact that the number of degrees of freedom these theories propagate is the same as those that general relativity does. 

In conclusion, the single valued functions that are unique up to canonical trnasforrmations (see following subsection) on the phase space co-ordinatised by $(g_{\mu\nu},\pi^{\mu\nu})$ that realise the constraint algebra \eqref{hc1}, \eqref{hc2}, \eqref{pb} are the constraints of general relativity \eqref{hamc}, \eqref{vc} in any number of dimensions.

The single valuedness of the Hamiltonian translates into the same for the RG flow it generator for the boundary theory, and this will be the case I restrict attention to in this article. It would be interesting to udnerstand what such a branched RG flow implies in a more general setting, which might be an interesting problem for future work to address. 
\subsection{Terms linear in the momenta}
One might ask wether a single trace beta beta function term which would appear as a term linear in momentum in the RG Hamiltonian is allowed by the consistency conditions. The answer is in the affirmative, if and only if the single trace beta function satisfies a certain gradient condition. This condition is exactly that which would allow such a term to be removed through a canonical transformation which would also redefine the potential. 
Suppose we started with the Hamiltonian 
$$H(\sigma)=\int \textrm{d}^{D}x  \sigma(x)\left(\frac{1}{\sqrt{g}}\left(\pi_{\mu\nu}\pi^{\mu\nu}-\frac{1}{D-1}\textrm{tr}\pi^{2}\right)+\beta_{\mu\nu}(g)\pi^{\mu\nu}+U(g)\right),$$
then it so happens that the consistency conditions demand that
\begin{equation}\beta_{\mu\nu}(g)=\frac{1}{\sqrt{g}}G_{\mu\nu\rho\sigma}\frac{\delta \mathcal{C}[g]}{\delta g_{\rho\sigma}},\label{bgr}\end{equation}
such that performing the canonical transformation 
\begin{equation}\pi^{\mu\nu}\rightarrow \pi^{\mu\nu}-\frac{\delta \mathcal{C}[g]}{\delta g_{\mu\nu}},\end{equation}
would remove the linear term and modify the potential as
$$U(g)\rightarrow U(g)-G^{\mu\nu\rho\sigma}\beta_{\mu\nu}\beta_{\rho\sigma},$$
where the beta function satisfy \eqref{bgr}. Of course, the consistency conditions also dictate the final form of the potential so that
\begin{equation}U(g)-G^{\mu\nu\rho\sigma}\beta_{\mu\nu}\beta_{\rho\sigma}=\sqrt{g}(R-2\Lambda).\end{equation}
To summarise, provided some $\mathcal{C}[g]$, the consistency conditions allow for a Hamiltonian of the form 
\begin{equation}H(\sigma)=\int \textrm{d}^{D}x  \sigma(x)\left(\frac{1}{\sqrt{g}}\left(\pi_{\mu\nu}\pi^{\mu\nu}-\frac{1}{D-1}\textrm{tr}\pi^{2}\right)+\beta_{\mu\nu}(g)\pi^{\mu\nu}-\sqrt{g}(R-2\Lambda)-G^{\mu\nu\rho\sigma}\beta_{\mu\nu}\beta_{\rho\sigma}\right),\end{equation}
where the beta function for the metric satisfies \eqref{bgr}. The fact that the terms linear in the momenta can be absorbed by a canonical transformation accompanied by a redefinition of the momenta was first noticed in the Lagrangian theory by Kuchar in \cite{Kuchar}. 
\subsection{Adding scalar matter}
If this truncation of single trace operators also includes a scalar single trace operator $O(x,\tau)$ with source $j(x,\tau)$ conjugate to the expectation value $\langle O(x,\tau)\rangle_{j,g}\equiv P(x,\tau).$ 
The presence of such an operator does modify the diffeomorphism and anomalous Weyl Ward identities, and hence the constraints on the dual phase space. The former now reads
$$\nabla_{\nu}\langle T^{\mu\nu}\rangle_{g,j}+\langle O\rangle_{g,j}\nabla^{\mu}j=0$$
\begin{equation}=> H_{\mu}(\xi^{\mu})=\int \textrm{d}^{D}x \sqrt{g} \xi^{\mu}(-\nabla_{\nu}\pi^{\nu}_{\mu}+P\partial_{\mu}j)=H^{g}_{\mu}(\xi^{\mu})+H^{j}_{\mu}(\xi^{\mu})=0.\end{equation}
The Poisson algebra of the above constraint remains unaltered. 

Then the fixing of the scalar Hamiltonian constraint requires a similar analysis of the previous section. Again, being agnostic to the form of the constraint and letting the algebra dictate what it shall eventually be. This leads to 
\begin{equation}H^{j}(\sigma)=\int \textrm{d}^{D}x\sigma(x)\left( G(j)P^{2}+\frac{1}{G(j)}\partial_{\mu}j\partial^{\mu}j+V(j)\right).\end{equation}
The analysis is fairly straightforward and a further simplification arises from the Poisson commutativity of the scalar and gravitational sectors. If we had many scalar operators, the above expression is modified into
\begin{equation}H^{j}(\sigma)=\int \textrm{d}^{D}x\sigma(x)\left( G_{AB}(j)P^{A}P^{B}+G^{AB}(j)\partial_{\mu}j_{A}\partial^{\mu}j_{B}+V(j)\right).\end{equation}
The total scalar constraint thus reads $H_{T}(\sigma)=H(\sigma)+H^{j}(\sigma),$ similarly for the momentum constraint $H_{T,\mu}(\xi^{\mu})=H_{\mu}(\xi^{\mu})+H^{j}_{\mu}(\xi^{\mu}).$ These satisfy the same constraint algebra as their constituents, as follows from the linearity of the Poisson brackets. 
Now if a linear in momentum term is added to the scalar Hamiltonian, i.e. if it is modified into 
$$H^{j}(\sigma)=\int \textrm{d}^{D}x\sigma(x)\left( G_{AB}(j)P^{A}P^{B}+P^{A}\beta_{A}(j)+G^{AB}(j)\partial_{\mu}j_{A}\partial^{\mu}j_{B}+V'(j)\right),$$
then the consistency conditions will demand here too that the beta function term is forced to be of the form
\begin{equation}\beta_{A}(j)=G_{AB}\frac{\delta \mathcal{C}[g,j]}{\delta j_{B}}.\end{equation}
The reason for using the same notation for this canonical transformation too is that it will also contribute towards the metric beta function given it's metric dependence. At the very least, if this is the integral of a purely scalar function of the sources $j_{A}$ and contains no derivatives, the integrand still will contain a factor of $\sqrt{g}$ and hence will not be ignored by the metric sector of the phase space. Thus the appropriate canonical transformation to consider is the one defined on the total phase space as
\begin{equation}
\left(\begin{array}{c} \pi^{\mu\nu} \\ P_{A} \end{array}\right) \rightarrow
\left(\begin{array}{c} \pi^{\mu\nu}-\delta \mathcal{C}/ \delta g_{\mu\nu}\\ P_{A}-\delta \mathcal{C}/\delta j^{A} \end{array}\right),
\end{equation})
and the effect of this on the total Hamiltonian constraint should be seen as requiring the redefinition of the potential 
$$
U(g)+V'(j)\rightarrow U(g)+V'(j)-G^{\mu\nu\rho\sigma}\beta_{\mu\nu}\beta_{\rho\sigma}-G^{AB}\beta_{A}\beta_{B},
$$
This should be the same as the original potential for the scalar field and gravity:
\begin{equation}U(g)+V'(j)-G^{\mu\nu\rho\sigma}\beta_{\mu\nu}\beta_{\rho\sigma}-G^{AB}\beta_{A}\beta_{B}=\sqrt{g}(R-2\Lambda)+V(j).\end{equation}
Now to study the vicinity of a fixed point of such holographic theories.

\section{Holographic Renormalization}
The previous sections dealt with aspects of the renormalization group flow which in principle will hold even if the theory under consideration is an effective field theory with a finite UV cutoff. This means that any solution to the flow equations cannot be extended to infinite flow time. The step of renormalization is to find a set of flows emanating from an ultraviolet fixed point so that the flow time can be taken to infinity. Corresponding to that there is a chart on the theory space with well defined transition maps corresponding to the renormalized sources. This perspective is emphasised in \cite{Liz}.

In order to define renormalized correlation functions and other such observables, one need only renormalize the generating functional. In the Hamiltonian system the flow gets mapped to, this corresponds to finding the on shell action defined with boundary conditions at large flow times. The on shell action is the same as Hamilton's principal functional so in order to study renormalization of the boundary theory, we need solve the Hamilton--Jacobi equation with large (radial) time asymptotics. This procedure is known as holographic renormalization (see for instance \cite{HRG}). The approach followed henceforth is that of Skenderis, Papadimitriou et. al. (see for instance \cite{Papa1}, \cite{Papa2}, \cite{PapaSken} and references therein) and is known as the Hamiltonian approach to holographic renormalization. 

Going back to the renormalized RG Hamiltonian \eqref{renham}, we notice that it is not a constraint, and its value is given by the integrated anomaly. Since the theory is renormalized and the cutoff is taken to infinity, the only terms appearing in this expression are those that remain finite in the UV limit. Restricting to the case where there are no marginal or relevant deformations, the only remaining source is the metric and the Hamiltonian itself is given by the trace of the renormalized energy momentum tensor: $H_{ren}=\langle T^{\mu}_{\mu}\rangle|_{\tau\rightarrow \infty}.$ The aim is to compute this in two and four dimensions from the more general unrenormalized RG Hamiltonian. 

From the bulk perspective, this implies fixing a specific set of asymtptoic boundary conditions. As is wel known, specifying the asymptotic boundary conditions also fixes the sign of the cosmological constant. Thus a parameter in the unrenormalized RG Hamiltonian is fixed though the demand that the flow remain meaninfgul at large values of $\tau$. 
In the coming section I will argue why the negative sign for the cosmological constant is the natural choice to turn the unrenormalized RG Hamiltonian into the renormalized one. 

\subsection{The sign of the Cosmological constant}

One can ask which canonical transformation removes the cosmological constant in favour of a term linear in the momenta. To determine the generating function for this we look at
\begin{equation*}2\Lambda-G_{\mu\nu\rho\sigma}\frac{\delta \mathcal{C}}{\delta g_{\mu\nu}}\frac{\delta \mathcal{C}}{\delta g_{\rho\sigma}}=0.\end{equation*}
If the cosmological constant is negative, the constant term in the potential can be rewritten in terms of the cosmological radius $l$: $2\Lambda=\frac{-(D(D-1))}{l^{2}}$. Then the above expression reads
\begin{equation*}
-\frac{D(D-1)}{l^{2}}=G_{\mu\nu\rho\sigma}\frac{\delta \mathcal{C}}{\delta g_{\mu\nu}}\frac{\delta \mathcal{C}}{\delta g_{\rho\sigma}}
\end{equation*}

Which is satisfied if 
\begin{equation}
\mathcal{C}[g]=\frac{D-1}{l}\int\textrm{d}^{D}x \sqrt{g}.
\end{equation}
In this case, the Hamiltonian constraint can be written as an equation for $\textrm{tr}\pi$:
\begin{equation}
\frac{2}{l}\textrm{tr}\pi=\frac{1}{\sqrt{g}}\left(\pi_{\mu\nu}\pi^{\mu\nu}-\frac{1}{D-1}\textrm{tr}\pi^{2}\right)-\sqrt{g}R. \label{trp}
\end{equation} 
Thus the Hamiltonian constraint can be turned into an equation for the trace of the conjugate momentum through a real canonical transformation in the case where the cosmological constant is negative. Given that the aim is the renormalized RG Hamiltonian at infinity whose left hand side is given by the trace of the enrgy momentum tensor, we see that such a canonical transformation is a step in the right direction. In order to fully appreciate this, one needs to study the Hamilton--Jacobi equations which shall be done in subsection 5.4 onwards. More evidence for this intuition can be gained by studying the dilatation operator at large times as well. This is the subject of the next subsection. 

\subsection{The Dilatation Operator}
I will present another reason why it is desirable to choose a negative cosmological constant if we want to tether the dynamics generated by the unrenormalized RG Hamiltonian to that of the renormalized one. Recalling that at the UV fixed point, the renormalized Hamiltonian is just the trace of the renormalized energy momentum tensor, as $\tau\rightarrow \infty$ we have
\begin{equation}
\partial_{\tau}(\cdot)\sim\left\{H^{\textrm{ren}},\cdot\right\}\propto\int \textrm{d}^{D}x \, \left(2g_{\mu\nu}\frac{\delta(\cdot)}{\delta g_{\mu\nu}}\right)\bigg|_{\tau\rightarrow \infty}. \label{dil}
\end{equation}
for later convenience I will denote $\delta_{g}(\cdot)\equiv2g_{\mu\nu}\frac{\delta(\cdot)}{\delta g_{\mu\nu}}$. 
So we see that the action of the Hamiltonian and that of the dilatation operator at infinity are thus identified in a simple manner.
This means that the renormalized Hamiltonian has a simple expression in terms of the asymptotic value of the metric tensor. The significance of the above expression \eqref{dil} is that it equates isolating the depedence of the metric and functions of it on $\tau$ to the problem of studying eigenfunctions of the dilataion operator. This is what the Hamiltonian approach to Holographic renormalization is based on. 
This identification corresponds to an asymptotic fixing of the Lagrange multipliers to $(\sigma,\xi^{\mu})=(1,0)$. 

Isolating the dependence of the metric itself in the large $\tau$ limit is easy to work out:
\begin{equation}\dot{g}_{\mu\nu}\stackrel{ \tau \rightarrow \infty }{ \mathlarger{\sim}}\delta_{g}g_{\mu\nu} = 2 g_{\mu\nu},\label{dilp}\end{equation}
and so $g_{\mu\nu}|_{\tau\rightarrow \infty}=e^{2 \tau}g_{(0)\mu\nu}.$ For future convenience it will help to rescale the exponent so that $g_{\mu\nu}=e^{\frac{2\tau}{l}}g_{(0)\mu\nu}$

This type of behaviour hints at asymptotic boundary conditions of the Anti-de Sitter type (assuming the radial direction is Euclidean). Additionally I will demonstrate more explicitly how the conformal algebra emerges at the fixed point which is nothing but the Lie bracket algebra of the Killing vectors that generate the asymptotic isometry group of Anti de Sitter space.

\subsection{The Conformal Algebra}
In the previous section, we see that when considering the U.V. fixed point of the theory the background metric takes the form $g_{\mu\nu}=e^{\frac{2\tau}{l}}g_{(0)\mu\nu}$ where $\tau\rightarrow \infty$. In this section, we will utilise this and focus on the case where $g_{(0)\mu\nu}=\eta_{\mu\nu}$, i.e. when the metric is conformally flat. Also, if we choose the the diffeomorphism transformations to be generated by 
\begin{equation}\xi^{\mu}=a^{\mu}+\omega^{\mu}_{\nu}x^{\nu}+\lambda x^{\mu}+2(b\cdot x)x^{\mu}-x^{2}b^{\mu},\end{equation}
and the Weyl factor given by $\frac{\partial\cdot \xi}{D}$, 
\begin{equation}\sigma(x)=\frac{\partial\cdot \xi}{D}=\lambda+2b\cdot x.\end{equation}
It then follows that the local RG transformations and diffeomorphisms generated by 
$$H(\lambda)+H_{\mu}(\lambda x^{\mu})=D^{(\lambda)},\, \,  H(2b\cdot x)+H(2b\cdot x x^{\mu}-x^{2}b^{\mu})=K^{(b)},$$

$$H_{\mu}(a^{\mu})=P^{(a)},\, \,  H_{\mu}(\omega^{\mu}_{\nu}x^{\nu})=J^{(\omega)}.$$
Then, it follows from the constraint algebra that these generators satisfy the following algebra:
\begin{eqnarray}
\left\{K^{(b)},D^{(\lambda)}\right\}=-K^{(\lambda b)}, \,\, \left\{P^{(a)},D^{(\lambda)}\right\}=P^{(\lambda a)}, \,\, \left\{K^{(b)},P^{(a)}\right\}=-D^{(a\cdot b)}+J^{(a\times b)},
\end{eqnarray}
$$ \left\{J^{(\omega)},K^{(b)}\right\} =K^{(\omega \cdot b)}, \,\,\left\{J^{(\omega)},P^{(a)}\right\} =P^{(\omega \cdot a)}, \,\, \left\{J^{(\omega)},J^{(\omega')}\right\} =J^{(\omega \cdot \omega')}.$$
Here $a\times b\equiv a^{[\mu}b^{\nu]}$, $\omega \cdot a\equiv \omega^{\mu}_{\nu}a^{\nu}$, $\omega \cdot b\equiv \omega^{\mu}_{\nu}b^{\nu}$  and $\omega \cdot \omega'\equiv \omega^{\mu}_{\nu}\omega'^{\nu}_{\rho}$. This is the conformal algebra that emerges as the residual symmetry group of the background geometry that arises at the UV fixed point. Note that in order to derive the above algebra, it is crucial that to note that the background given by the conformally flat metric with the infinite Weyl factor, the Hamiltonian constraint commutes strongly. This is because the structure function vanishes $e^{-\frac{2\tau}{l}}g_{(0)}^{\mu\nu}(\sigma\partial_{\nu}\sigma'-\sigma'\partial_{\nu}\sigma)\rightarrow 0$ as $\tau \rightarrow\infty$. This is consistent with the fact that Weyl transformations commute.

\subsection{Hamilton--Jacobi Equations}
The Hamilton--Jacobi equation is obtained from setting the momenta equal to derivatives of Hamilton's principal functional:
\begin{eqnarray}\pi^{\mu\nu}=\frac{\delta \mathcal{S}}{\delta g_{\mu\nu}}, \ P^{A}=\frac{\delta \mathcal{S}}{\delta j_{A}}.\end{eqnarray}
This is nothing but a restatement of \eqref{orig} making the identification of the generating functional with Hamilton's principal functional. The Hamilton--Jacobi equations then take the form 
\begin{eqnarray}
G_{\mu\nu\rho\sigma}\frac{\delta \mathcal{S}}{\delta g_{\mu\nu}}\frac{\delta \mathcal{S}}{\delta g_{\rho\sigma}}+G_{AB}\frac{\delta \mathcal{S}}{\delta j_{A}}\frac{\delta \mathcal{S}}{\delta j_{B}}-\sqrt{g}(R-2\Lambda)+V(j)=0, \\
\nabla_{\nu}\frac{\delta \mathcal{S}}{\delta g_{\mu\nu}}+(\nabla^{\mu}j^{A})\frac{\delta \mathcal{S}}{\delta j_{A}}=0.
\end{eqnarray}
These are nothing but a restatement of the local renormalization group equations satisfying the holographic Wess--Zumino consistency conditions. 

The Hamilton--Jacobi equations can also be seen as a canonical transformation away from zero momenta. From this perspective, the beta functions are given by the gradient formula where $\mathcal{C}$ is replaced by $\mathcal{S}$. 
The fixed point of interest  where the trace anomaly shall be computed is characterised by the zero value of the scalar beta function, so Hamilton's principal function will have trivial $j_{A}$ dependence and the scalar potential $V(j)$ takes a constant value. It adds to the cosmological constant term in the gravitational Hamiltonian. Thus in order to study this fixed point, it will suffice to consider pure gravity. 
\subsection{Solving the Hamilton--Jacobi Equations}
The approach followed in this subsection and the next summarises that which is described in detail in \cite{Papa1}, \cite{Papa2} and introduced in \cite{PapaSken}. As mentioned before, the idea is to expanding Hamilton's principal functional in eigenfunctions of $\delta_{g}$:
\begin{eqnarray}
\mathcal{S}=\mathcal{S}_{(0)}+\mathcal{S}_{(1)}+\mathcal{S}_{(2)}+\cdots, \ \mathcal{S}_{(k)}=\int \textrm{d}^{D}x \sqrt{g} \mathcal{L}_{(k)},
\end{eqnarray}
such that these functions satisfy
\begin{equation}\delta_{g}\mathcal{S}_{(k)}=(D-k)\mathcal{S}_{(k)}. \label{ds}.\end{equation}
This aids us with determining the expansion of the momentum:
\begin{equation}\pi^{\mu\nu}=\pi^{\mu\nu}_{(0)}+\pi^{\mu\nu}_{(1)}+\pi^{\mu\nu}_{(2)}+\cdots, \end{equation}
through the realtion 
So the relation \eqref{ds} can be written through the momenta as
\begin{equation}\textrm{tr}\pi_{(k)}=(D-k)\mathcal{L}_{(k)},\end{equation}
and recalling that one can define the momentum through the functional derivative of the principal function with respect to the metric. Recalling the canonical transformation performed to trade the negative cosmological constant for a term linear the momentum, we see already that $\mathcal{S}$ can already eb decomposed as
\begin{equation*}
\mathcal{S}=\frac{D-1}{l}\int \textrm{d}^{D}x \sqrt{g}+ \mathcal{S}',
\end{equation*}
where the expansion $\mathcal{S}'$ goes from order 1 onwards. Then the equation \eqref{trp} which is equivalent to the Hamilton--Jacobi equations reads
\begin{equation}
\frac{2g_{\mu\nu}}{l}\frac{\delta\mathcal{S}'}{\delta g_{\mu\nu}}=\frac{1}{\sqrt{g}}G_{\mu\nu\rho\sigma}\frac{\delta \mathcal{S}'}{\delta g_{\mu\nu}}\frac{\delta \mathcal{S}'}{\delta g_{\rho\sigma}}-\sqrt{g}R.
\end{equation}
Notice that the left hand side is the generator if dilatations with Weyl factor $e^{\frac{2}{l}}$. This can be seen as follows
\begin{equation*}
2g_{\mu\nu}\frac{\delta \mathcal{S}'}{\delta g_{\mu\nu}}=\frac{\delta \mathcal{S}'(e^{2\phi(x)}g)}{\delta \phi(x)}\bigg|_{\phi=0}.
\end{equation*}
Then recalling the relation \eqref{dilp}, we see that this can be turned into the following equation:
\begin{equation*}
2\partial_{\tau}\mathcal{S}'(e^{\frac{2\tau}{l}}g_{(0)})=\int \textrm{d}^{D}x 
\left(\frac{1}{\sqrt{g}}G_{\mu\nu\rho\sigma}\frac{\delta \mathcal{S}'}{\delta g_{\mu\nu}}\frac{\delta\mathcal{S}'}{\delta g_{\rho\sigma}}\right)\bigg|_{g=\exp{\left(\frac{2\tau}{l}\right)}g_{(0)}}-e^{(D-2)\frac{2\tau}{l}}R(g_{(0)})\sqrt{g_{(0)}}.\end{equation*}
then expanding the term in the prantheses in terms of the eigenvalues of the dilatation operator:
\begin{equation*}
\left(\frac{1}{\sqrt{g}}G_{\mu\nu\rho\sigma}\frac{\delta \mathcal{S}'}{\delta g_{\mu\nu}}\frac{\delta\mathcal{S}'}{\delta g_{\rho\sigma}}\right)\bigg|_{g=\exp{\left(\frac{2\tau}{l}\right)}g_{(0)}}=\sum^{\infty}_{m,n=1}e^{\frac{2\tau}{l}(-D+2m+2n)}\frac{1}{\sqrt{g_{(0)}}}G_{\mu\nu\rho\sigma}\frac{\delta \mathcal{S}_{(m)}}{\delta g_{\mu\nu}}\frac{\delta \mathcal{S}_{(n)}}{\delta g_{\rho\sigma}}.
\end{equation*}
Plugging this back in the above expression, you get
\begin{equation}
\sum^{\infty}_{n=1}\frac{(D-k)}{l}\mathcal{S}_{k}=\int \textrm{d}^{D}x e^{\frac{2\tau}{l}(2-D)}\sqrt{g_{0}}R+\sum^{\infty}_{m,n=1}\int \textrm{d}^{D}x e^{\frac{2\tau}{l}(-D+2m+2n)}\frac{1}{\sqrt{g_{(0)}}}G_{\mu\nu\rho\sigma}\frac{\delta \mathcal{S}_{(m)}}{\delta g_{\mu\nu}}\frac{\delta \mathcal{S}_{(n)}}{\delta g_{\rho\sigma}}.
\end{equation}
By comparing powers of $e^{\frac{2\tau}{l}}$, one finds that 
\begin{equation}
\mathcal{S}_{(1)}=\frac{l}{2(D-2)}\int \textrm{d}^{D}x \sqrt{g_{(0)}}R(g_{(0)}).
\end{equation}
Then, there is a set of recursion relations for the higher order terms
\begin{equation}
2(D-k)\mathcal{S}_{(k)}=\sum^{k-1}_{n=1}\int \textrm{d}^{D}x \frac{1}{\sqrt{g}_{(0)}}G_{\mu\nu\rho\sigma}\pi^{\mu\nu}_{(n)}\pi^{\rho\sigma}_{(k-n)}. \label{recur}
\end{equation}
The terms in the expansion of the momenta have been inserted for notational ease. 
\subsection{Pole Terms and the Conformal Anomaly}
Clearly, in various even dimensions, the expansion of $\mathcal{S}$ goes only up to $k=D$ due to the presence of poles. 
Let us start by lookig at the $D=2$ case. Here, $\mathcal{S}_{(1)}$ is divergent. To cure this, dimensional regularisation is employed, as mentioned before to set $\tau_{0}=\frac{1}{D-2},$ and then one defines
$$\mathcal{L}_{(2)}|_{\tau_{0}}=-\frac{2}{l}\tau_{0}\tilde{\mathcal{L}}_{(2)}|_{\tau_{0}},$$
so that
\begin{equation}\tilde{\mathcal{L}}_{(2)}=-\frac{1}{2}\textrm{tr}\pi_{(2)}.\end{equation}
This is finite as $\tau\rightarrow \infty$ and in that limit this defines the trace Ward identity, again from recalling the relation $\lim_{\tau\rightarrow \infty}\textrm{tr}\pi_{(2)}=\langle T^{\mu}_{\mu}\rangle_{ren}$. So, thus we find the expression for the anomaly at the fixed point:
\begin{equation}H^{Ren}(g_{(0)})=\langle T^{\mu}_{\mu}\rangle_{ren}=l\int \textrm{d}^{D}x \sqrt{g_{(0)}}\left(R(g_{(0)})\right).\end{equation}
Reinstating units, one sees that the well known relationship $c\propto \frac{l}{G}$ is recovered. Then, doing the same for the pole term in $D=4$ one finds:
\begin{equation*}\textrm{tr}\pi_{(4)}=G_{\mu\nu\rho\sigma}\pi^{\mu\nu}_{(2)}\pi^{\rho\sigma}_{(2)}\end{equation*}
from which it follows that:
\begin{equation} H^{ren}(g_{(0)})=\langle T^{\mu}_{\mu}\rangle_{ren}=\left(\frac{1}{3}R^{2}-R^{\mu\nu}R_{\mu\nu}\right).\end{equation}

This is known as the Holographic anomaly, which is one of the celebrated results of the AdS/CFT correspondence: \cite{Skenderis}, \cite{Imbimbo1}.
Comparing this to the general expression for the anomaly in conformal field theories in four dimensions, i.e.
$$\mathcal{A}=\left(\frac{c}{3}-a\right)R^{2}+(-2c+4a)R^{\mu\nu}R_{\mu\nu}+(a-c)R^{\mu\nu\rho\sigma}R_{\mu\nu\rho\sigma},$$

we see that the relation $a=c$ between the anomaly coefficients is implied by the above expression. 
Another way to see how the $a=c$ condition arises in the local holographic RG was demonstrated by Nakayama in \cite{Nakayama1}.

\section{Conclusion}
In this article, the case is made for the Holographic Wess--Zumino consistency condition is a defining feature of the quantum field theories possessing local gravity duals. This consistency condition encodes the emergence of diffeomorphism invariance in the holographically dual bulk. It encodes the response of holographic field theories to local renormalization group transformations such that the energy scale or RG flow time that is identified with the emergent direction is treated on equal footing with the other directions the field theory inhabits. 

So, this condition has to be reconcilable with the condition that the $a$ and $c$ anomaly coefficients are equated at the fixed point for conformal field theories possessing General Relativity duals. This fact follows from the computation of the holographic anomaly, which in turn is derived through solving Einstein Hamilton--Jacobi equations in asymptotically (A)dS spaces. The Hamiltonian underlying the latter has a structure fixed completely by the HWZ conditions. Running this logic backwards is how the consistency conditions lead to the computation of the holographic anomaly. 

\section*{Acknowledgments}
I would like to thank Sung-Sik Lee for fruitful discussions regarding the quantum renormalization group. I would also like to thank Aldo Riello, Bianca Dittrich, Henrique Gomes and Lee Smolin for discussions and support for this work.

This research was supported in part by Perimeter Institute for Theoretical Physics as well as by grant from NSERC and the John Templeton Foundation. Research at
Perimeter Institute is supported by the Government of Canada through the Department of Innovation,
Science and Economic Development Canada and by the Province of Ontario through the Ministry of
Research, Innovation and Science

\end{document}